# Extraordinary acoustic reflection enhancement by acoustically-transparent thin plates

Zhaojian He,[1] Shasha Peng,[1] Chunyin Qiu,[1*] Rui Hao,[1] Gangqiang Liu,[1] Manzhu Ke,[1] Jun Mei,[2] Weijia Wen,[2] and Zhengyou Liu[1*]

[1]Key Laboratory of Artificial Micro- and Nano-structures of Ministry of Education and School of Physics and Technology, Wuhan University, Wuhan 430072, China

[2]Department of Physics, Hong Kong University of Science and Technology, Clear Water Bay, Kowloon, Hong Kong, China

**Abstract**

We report an observation of the extraordinary high reflection of acoustic waves in water by thin epoxy plates partitioned by subwavelength cuts, whereas such plates without structure are acoustically-transparent as the acoustic properties of epoxy are close to water. It is demonstrated that this exotic phenomenon results from the resonant excitation of the local modes within the individual pieces derived by the cuts. The experiment agrees well with the theory.

---

[*] Corresponding authors. cyqiu@whu.edu.cn; zyliu@whu.edu.cn

Recently, the extraordinary transmission enhancement of light through metallic plates perforated with periodical subwavelength apertures has attracted much attention [1-3], because of its prospective applications. The physical origin of the effect has been widely attributed to the resonant excitation of surface plasmon on metal surfaces, but there is still a debate on the relative contribution of the surface wave versus the diffraction mechanism [4-10]. Similar transmission phenomena of acoustic waves through solid plates drilled with subwavelength apertures have also been reported [10-15], which is attributed to either the Fabry-Pérot (FP) resonance in the holes/slits, or the coherent diffraction induced by the periodicity. In most of these studies, rigid plates were employed or assumed, hence it is the apertures, either individually or collectively, that account for the phenomena. With the elasticity of the solid plates fully taken into account, the acoustic transmission enhancement originated in the resonant excitation of the intrinsic Lamb waves [16,17] has been observed too.

For a soft plate (e.g. PMMA or epoxy plate) immersed in water, one may think by intuition that it is transparent for acoustic waves, since the acoustic impedance of the plate is close to water. Recently, however, we have demonstrated by simulation that for a thick PMMA plate drilled with periodical holes, extraordinary reflections due to the FP resonance in the holes, and transmission minima due to the resonant excitation of the Stoneley waves on the plate surfaces can be observed [18]. In this Letter, we will further show by both the simulation and experiment that even for a very thin (with respect to the wavelength considered) epoxy plate that can not sustain any FP resonance, the acoustic reflection enhancement (ARE) can still be achieved when it is structured by subwavelength cuts. It is found that this extraordinary ARE phenomenon arises from the resonant excitation of the local modes within the single piece cut from the plate, rather than due to the collective diffraction effect as in the case of rigid plates [11-14]. Throughout this Letter, a finite-element analysis and solver software package (Comsol Mutiphysics 4.0) is adopted to conduct the theoretical simulations, and the ultrasonic technique is employed to measure the transmission spectra. The experimental results agree well with the theoretic ones.

Our samples are the thin epoxy plates (density $\rho = 1.8 g/cm^3$, longitudinal and transverse wave velocities $c_l = 2.74 km/s$ and $c_t = 1.6 km/s$) of thickness $t = 0.4 mm$, with or without structures, immersed in water ($\rho = 1.0 g/cm^3$ and $c_l = 1.49 km/s$). A sample with periodical subwavelength cuts along $x$ direction is schematically illustrated in Fig. 1(a). These cuts partition the plate into separate strips of width $w = 1.8 mm$ and period $a = 2.3 mm$. By simulating a plane wave incident onto the periodical structure and transmitted through it, transmission coefficient is extracted by the finite-element method. The numerical results (black curves) are presented in Figs. 1(b) and 1(c) respectively for the incident angles $\theta = 0^o$ and $\theta = 10^o$, which are both featured, beyond our expectation, by a remarkable dip at water wavelength $\lambda \simeq 2.8 mm$, indicating a very high reflection. For comparison, the transmission spectra for the unstructured plate with the same thickness (blue curves) are also provided in the figures, in which good transparency is exhibited as the acoustic impedance of epoxy matches water well. High transmission of the reference plate also attributes to the fact that the wavelength considered here is much larger than the plate thickness, which also account for no FP resonant peaks observed in the spectra. To verify the extraordinary reflection, experimental measurement based on the ultrasonic transmission technique [16,17] has been performed. In the experiment, the sample is clamped with desired tilted angles and placed in between a couple of generating and receiving transducers with diameters of $25.4 mm$ and central frequencies at $0.5 MHz$. The whole assembly is immersed in a big water tank. The measured transmission spectra (red curves) in Figs. 1(b) and 1(c) reproduce satisfactorily the extraordinary ARE phenomena [19].

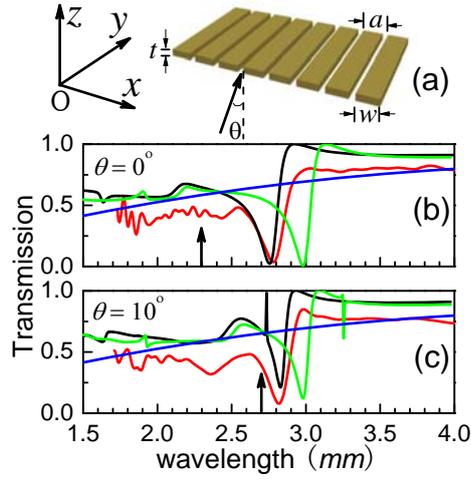

FIG. 1 (Color Online). (a) Schematic illustration for the structured sample consisting of a thin epoxy plate partitioned by subwavelength cuts translated along *x* direction. The plate thickness is $t = 0.4mm$, the width of the strips derived by the cuts is $w = 1.8mm$, and the translational period is $a = 2.3mm$. (b) The measured (red) and calculated (black) transmissions at an angle $\theta = 0^o$ incident onto the structured sample, compared with that of the reference plate (blue), where the horizontal-axis represents the wavelength in water. (c) The same as (b), but for $\theta = 10^o$. In addition, in (b) and (c) we also provide the numerical transmissions (green) for the sample with identical geometry except for replacing the strip width by $2.0mm$. Here the arrows mark the frequencies of Wood anomalies for the corresponding incident angles.

In view of the weak angular dependence of the dip position, the ARE observed here should not arise from the diffraction effect by the periodical sample. This is further validated numerically by the sensitive red shift of the transmission dip as the strip width is increased to $2.0mm$ with keeping the period invariant, as shown by the green curves in Fig. 1. In fact, because of the good transparency of the sample, acoustic waves can mostly transmit through it in a direct way, and hence the diffraction effect due to the periodicity should be inconspicuous in the spectra. Just as it is, the Wood anomaly denoted by the arrows, a typical diffraction effect is indeed not obvious in Fig. 1. This is in contrast to the case of rigid plate with structures, in which, due to diffraction, the Wood anomaly, as well as the adjoined nearly-total transmission peak on the longer wavelength side (this enhanced transmission is the focus of the extensive investigations recently) can always be clearly manifested [11-14]. In order to understand the physical origin of the extraordinary ARE effect, we study the field distribution of acoustic waves inside and around the sample with a finite-element simulation at the wavelength corresponding to the transmission dip. Figure 2 shows the field distribution excited by a plane wave incident normally from below. One observes that the field is mainly distributed around the strip, and in the strip it is weak in the middle but strong at the edges, indicating a contracting/stretching vibration. This suggests that the resonance is connected with the local mode of a single strip, and the coherent resonances of all strips give rise to the high reflection of acoustic waves.

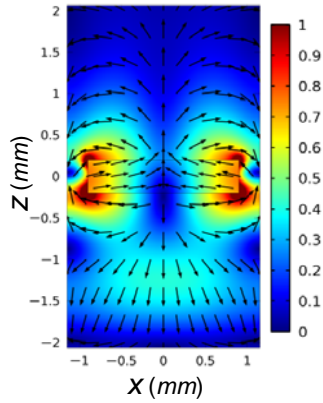

FIG. 2 (Color Online). Cross section view of the field distribution inside and around one strip (marked by the dashed white frame) of the structured sample, incident normally by a plane wave at the resonant wavelength $\lambda \simeq 2.8mm$, where the color scale displays the field amplitude and arrows indicate the field direction.

To confirm the above conjecture, we study the effect of a single epoxy strip. By launching a Gaussian acoustic beam of width $10mm$ onto a single strip of width $1.8mm$, the transmission is extracted. As shown in Fig. 3(a), the transmission spectrum exhibits a remarkable dip near the wavelength $\lambda \simeq 2.5mm$, close to the case with all strips present ($\lambda \simeq 2.8mm$). Figure 3(b) provides the field distribution around the strip at this particular wavelength, which displays a strong field at the edges of the strip, accompanied by a weak field in the middle, similar to the all-strips case in Fig. 2.

It is noted that there is a slight shift of the transmission dip toward the longer wavelength for the all-strips case compared with the single-strip situation. This stems from the coupling of local modes in adjoining strips, as to be interpreted in the following.

The resonant modes in a thin strip are essentially the standing wave modes of the Lamb waves in the plate extending over the strip breadth. For an infinite solid plate immersed in liquid, there are rich numbers of intrinsic modes associated with the plate [16-18,20], generally classified as the symmetric and antisymmetric types. Specifically for a thin epoxy plate in water, only the zero-order symmetric and antisymmetric Lamb waves play parts (the higher-order ones have lower-cutoff frequencies and are out of the scope considered here). Figure 3(c) shows the dispersion relations of the two modes for an epoxy plate of thickness $0.4mm$, solved by including the full elastic equations [20], where the horizontal-axis represents the wavelength $\lambda$ in water (corresponding to frequency) and vertical-axis denotes the wavelength $\lambda_s$ for the plate modes. The dispersion curve of the symmetric modes lies above the water line, and for this type of modes the planar stretching/contracting vibration is dominant in the plate, while the dispersion curve of the antisymmetric modes lies below the water line, and for that type of modes the bending vibration is dominant in the plate. Besides the (nonleaky) antisymmetric Lamb modes, the (leaky) symmetric ones can not be excited at small angles from water as well, so they do not manifest themselves in the transmission spectra of the uniform plate (see blue curves in Fig. 1). Standing wave modes of the Lamb waves can be created in a strip owing to the confinement of the strip in width. The fundamental symmetric standing Lamb modes (SSLMs) can be easily excited by the incident waves since they are connected with the compression/expansion motion, well compatible with the incident waves; whereas those antisymmetric ones can hardly be excited, since they are connected with the bending of the strip or shear motion inside the strip, badly compatible with the incident waves [19]. Therefore, the pounced transmission dips observed in Fig. 1 stem from the resonant excitation of the SSLMs in a single strip. The fundamental

SSLM extends a half-wavelength across the strip breadth plus a correction by the strip thickness, i.e., $\lambda_s/2 = w + t = 2.2mm$, which gives directly the resonant frequency or wavelength from the dispersion curve as indicated by the arrow in Fig. 3(c). We see that there is an excellent agreement of the estimation with that revealed in the transmission [see Fig. 3(a)]. For more strips as in the all-strips case, the in-phase stretching/contracting motion of the neighboring strips slows down the vibration of each strip, resulting in a red shift of the SSLM. If the separation between strips is increased, the SSLMs within individual strips would be less influenced and the transmission dip would coincide with that for the single strip case. To confirm it, calculation and experiment are carried out for a system similar to Fig. 1(a), but with a bigger period $a = 4.3mm$. As displayed in Fig. 4, the resonant dip for this system indeed approaches the wavelength $\lambda = 2.5mm$. Here the measured transmission dip is much shallower than the theoretic case, which is mostly due to the finite size effect in experiment: the incident acoustic wave covers only six periods of the sample.

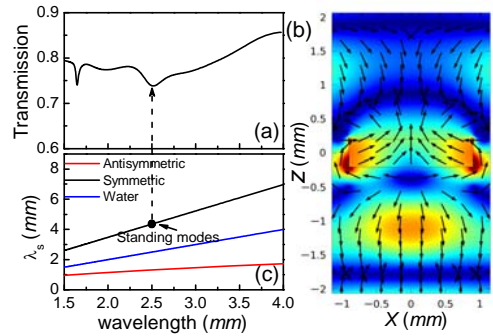

FIG. 3 (Color Online). (a) The calculated transmission of a Gaussian beam passing through a single epoxy strip of width $1.8mm$. (b) The field distribution inside and around the strip at the resonant wavelength $\lambda \simeq 2.5mm$. (c) The dispersion curves of the symmetric (black) and antisymmetric (red) Lamb waves in a flat plate of thickness $0.4mm$, reference to that of acoustic waves in water (blue). The arrow points to the position of the fundamental standing wave mode in the strip.

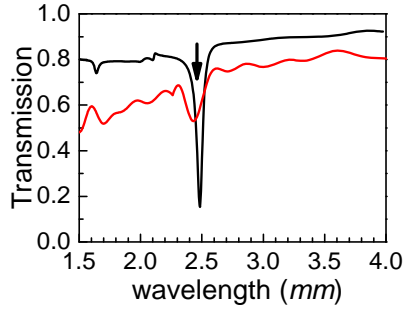

FIG. 4 (Color Online). The numerical (black) and experimental (red) transmission spectra at normal incidence for a structured sample similar to Fig. 1(a) but with a bigger period $a = 4.3 mm$.

It is straightforward to infer that the extraordinary ARE effect can also be observed for the thin plate partitioned by parallel cuts translated along two, e.g. perpendicular directions. Fig. 5(a) illustrates the epoxy plate after applying the parallel cuts translated along the *y* direction, following the parallel cuts translated along the *x* direction [see Fig. 1(a)]. By the two parallel cuts, the plate is partitioned into periodically distributed separated square tiles. Each tile has a width of $1.8mm$ and the translation period is $2.3mm$. The transmission spectrum of acoustic waves for this structure is shown in Fig. 5(b), which exhibits a resonant dip at wavelength $\lambda \simeq 2.5mm$. Figure 5(c) gives the field distribution inside and around a tile at resonance, revealing again a character of SSLMs, but now in a two dimensional region. The field is weak in the middle but strong along the edges especially in the four corners. This feature of the field distribution allows one to fabricate such structure practical by attaching each tile at the center to the intersections of a rigid grid, as depicted in the Fig. 5(d).

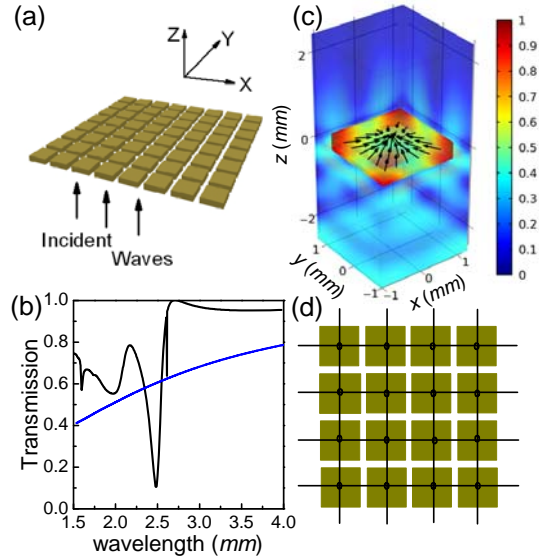

FIG. 5. (Color Online) (a) The thin epoxy plate sample partitioned by translational parallel cuts along *x* and *y* directions, where the side length of the square tile derived by the cuts is $1.8mm$, and the translational periods are both $2.3mm$. (b) The calculated transmission spectra for the structured sample (black) and the reference plate (blue). (c) The field distributions in a cross section through the center of a tile and perpendicular to the *y* axis at resonance. (d) A schematic illustration to fabricate the sample in practice.

In conclusion, we have observed a remarkable ARE effect in the thin epoxy plate partitioned by translational subwavelength cuts. By analyzing in detail the properties of vibration field at high reflection, we find that this extraordinary phenomenon is caused by the resonant response of the individual pieces. The standing wave modes of the symmetric Lamb waves in the pieces account for the resonant responses. It is worth noting that such local modes stem from the complex interaction between the longitudinal and transverse waves, which have no optic counterpart. One-dimensional (with the partitioned pieces being strips) and two-dimensional (with the partitioned pieces being tiles) cases are both discussed. Application exploitations can be stimulated by the phenomena reported here, such as in blocking sound or creating strong local-field.

**Acknowledgement:**

This work is supported by the National Natural Science Foundation of China (Grant Nos. 10874131, 50425206 and 10974147); Hubei Provincial Natural Science Foundation of China (Grant No. 2009CDA151); and NSFC/RGC joint research grants 10731160613 and N_HKUST632/07.